\documentclass[aps,twocolumn]{revtex4}
\usepackage{ulem}
\usepackage{times}
\usepackage{graphicx}
\usepackage{amssymb}

\begin{document}

\title{Isospin Constraints on the Parametric Coupling Model for Nuclear Matter}
\author{Guilherme F. Marranghello}
\affiliation{Centro de Ci\^encias Exatas e Tecnol\'ogicas \\
Universidade Federal do Pampa, Bag\'e - RS, Brazil}
\author{Constan\c ca Provid\^encia, Alexandre M. S. Santos}
\affiliation{Centro de F\'{\i}sica Computacional - Department of Physics\\
University of Coimbra - P-3004 - 516 - Coimbra - Portugal}

\begin{abstract}
We make use of isospin constraints to study the parametric coupling model and the properties of asymmetric nuclear matter. 
Besides the usual constraints for nuclear matter - effective nucleon mass and the incompressibility at saturation density 
- and the neutron star constraints - maximum mass and radius - we have studied the properties related with the symmetry energy. 
These properties have constrained to a small range the parameters of the model.
We have applied our results to study the thermodynamic instabilities in the liquid-gas phase transition  as well as
the neutron star configurations.
\end{abstract}

\maketitle

\section{Introduction}

The search for an effective nuclear matter theory has been one of the main goals of nuclear physics (see, for instance, the pioneering work of Fetter and Walecka \cite{Fetter_71}). The advances in experimental data acquisition has played an essential role on the improvement of nuclear models. One of the most required parameters to constrain symmetric nuclear matter models, the incompressibility \cite {blaizot}, has had many contributions since the non-linear Boguta and Bodmer model \cite{BB} was proposed as  an alternative to the Walecka model \cite{wal}.
The derivative coupling models proposed by Zimanyi and Moszkowski were also inspired to correct distortions in the incompressibility and effective nucleon mass determinations \cite{ZM}. 

In this work we make use of the so called parametric coupling model
(PCM) \cite{taurines} which describes nuclear matter with
density dependent coupling constants. The PCM, presented in the next
section has appeared in the literature after a first attempt to unify
Walecka and ZM models in an unique model \cite{delfino}. The PCM was
successful in reproducing the results obtained in the Walecka
\cite{wal}, ZM \cite{ZM} and exponential coupling models \cite{koepf}
and even more successful in reproducing nuclear matter and neutron
star main properties using special choices for the free parameters that define the density dependence of
the coupling constants. The PCM has been used to study a wide range of problems, specially the nuclear matter compression modulus \cite{veronica}, the inclusion of strange meson fields \cite{razeira} and the hadron-quark phase transition \cite{marranghello,grohmann}. 

Presently, an important issue in nuclear physics is to constrain the nuclear matter equation of state (EOS) from compact star properties or from laboratory measurements such as the ones planned for the next-generation of exotic radioactive beam facilities at CSR/China, FAIR/Germany, RIKEN/Japan, SPIRAL2(GANIL)/France or the planned Facility of Rare Isotope Beams/USA, where the high density behavior of the symmetry energy will be further studied experimentally. While saturation properties of symmetric nuclear matter such as the saturation density, binding 
energy and incompressibility are quite well settled, properties of asymmetric nuclear matter, such as 
the density dependence of the symmetry energy are much  less constrained.

Many authors have recently tested nuclear models, both phenomenological \cite{brown00,typel01,horo01,chen05,stei05,li08,cente09,chen09,pieka09,danie09,JunXu_09}, variational \cite{stei05} or microscopic \cite{li08,isaac09} such as the Brueckner-Hartree-Fock formalism, in order to impose constraints or verify their compatibility with data coming from heavy ion collisions, giant monopole resonances or isobaric analog states. In particular, recent experimental constraints from isospin diffusion in heavy-ion collisions (HIC) give for the slope parameter of the symmetry energy the value L = 88$\pm$25 MeV and for the isospin incompressibility coefficient $K_\tau=-500\pm 50$ MeV \cite{li08,chen05}. The latter is in agreement with the value of $K_\tau = -550\pm 100$ MeV predicted by the independent measurement of the isotopic dependence of the giant monopole resonance  in Sn isotopes \cite{garg07,li07}  and the value $K_\tau=-500 \small{\begin{array}{c} +125\\-100\end{array}}
$ MeV obtained from the study of neutron skins \cite{cente09}.
On the other hand isoscaling in HIC favors $L \sim 65$ MeV \cite{shetty07}  and nucleon emission ratios favor $L \sim  55$ MeV \cite{famiano06}. 

The main goal of the present work is to constrain even more the free parameters
in the PCM by studying  the symmetry energy and its density dependence
through its slope and its second derivative with respect to the density. We try to establish some correlations between the symmetry energy and its density derivatives with quantities with astrophysical interest such as the crust-core crossing density, the density at muon and at strangeness onset, the maximum star mass, and the threshold star mass for direct Urca \cite{urca}.

 In section \ref{sec:PCM} we summarize the PCM, in section \ref{sec:esym} the symmetry energy dependence on the density is discussed, the thermodynamic instabilities are discussed in section \ref{sec:thermo} and the neutron star properties in section \ref{sec:ns}. Finally we draw some conclusions in section \ref{sec:concl}.

\section{Parametric Coupling Model\label{sec:PCM}}

We first study the nuclear matter properties within the  PCM. This model is based on the  Walecka model \cite{wal}. However, in order to correct the values of the effective nucleon mass and incompressibility of symmetric nuclear matter at saturation density, Taurines et. al. \cite{taurines} have introduced  parametric couplings. They are  introduced in a similar way as Zimanyi and Moszkowski included the derivative couplings in their models. Here we briefly describe the model and the reader is addressed to  Ref. \cite{taurines} for further details.
The PCM Lagrangian density reads
\begin{eqnarray} {\cal{L}} & = &
 \sum\limits_{B} \bar{\psi}_B [i \gamma_\mu \partial^\mu - (M_B-
g_{\sigma B}^\star \sigma)-g_{\omega  B}^\star \gamma_\mu
\omega^\mu ]\psi_B \nonumber \\ & - & \sum\limits_{B} \psi_B
[\frac12 g_{\rho B}^\star \gamma_\mu
\mbox{\boldmath$\tau$}\cdot \mbox{\boldmath$\rho$}^\mu] \psi_B
 +  \sum\limits_\lambda \bar{\psi}_\lambda [i \gamma_\mu \partial^\mu -
m_\lambda] \psi_\lambda \nonumber \\ & + & \frac12(\partial_\mu
\sigma \partial^\mu \sigma - {m_\sigma}^2 \sigma^2) - \frac14
\omega_{\mu \nu} \omega^{\mu \nu} + \frac12 {m_\omega}^2
\omega_\mu \omega^\mu \nonumber \\ & - & \frac14
\mbox{\boldmath$\rho$}_{\mu \nu} \cdot
\mbox{\boldmath$\rho$}^{\mu \nu} + \frac12m_\rho^2
\mbox{\boldmath$\rho$}_\mu \cdot \mbox{\boldmath$\rho$}^\mu
\nonumber \\
\end{eqnarray}
where
\begin{equation}
g_{\sigma B}^\star \equiv m^\star_{\alpha B} g_{\sigma}  \, \, ;
\, \, g_{\omega B}^\star \equiv m^\star_{\beta B} g_{\omega}  \,
\, ; \, \, g_{\rho B}^\star = m^\star_{\gamma B} g_{\rho} \,
\end{equation}
and
\begin{equation}
m^\star_{n B} \equiv (1 + \frac{g_{\sigma} \sigma}{n M_B})^{-n} \,
; \, \, n = \alpha, \beta, \gamma \, \label{gefome1}.
\end{equation}
In the equations above, $\psi_B$ represents the baryon fields that can
be summed over the whole baryon octet. The baryon fields are coupled
to the meson fields $\sigma$, $\omega$ and $\rho$. The electron and
muon fields appear as $\psi_\lambda$ and must be introduced for the
description of stellar matter.
\bigskip
\begin{figure}[!h]
\begin{center}
\begin{tabular}{c}
\includegraphics[width=0.8\linewidth]{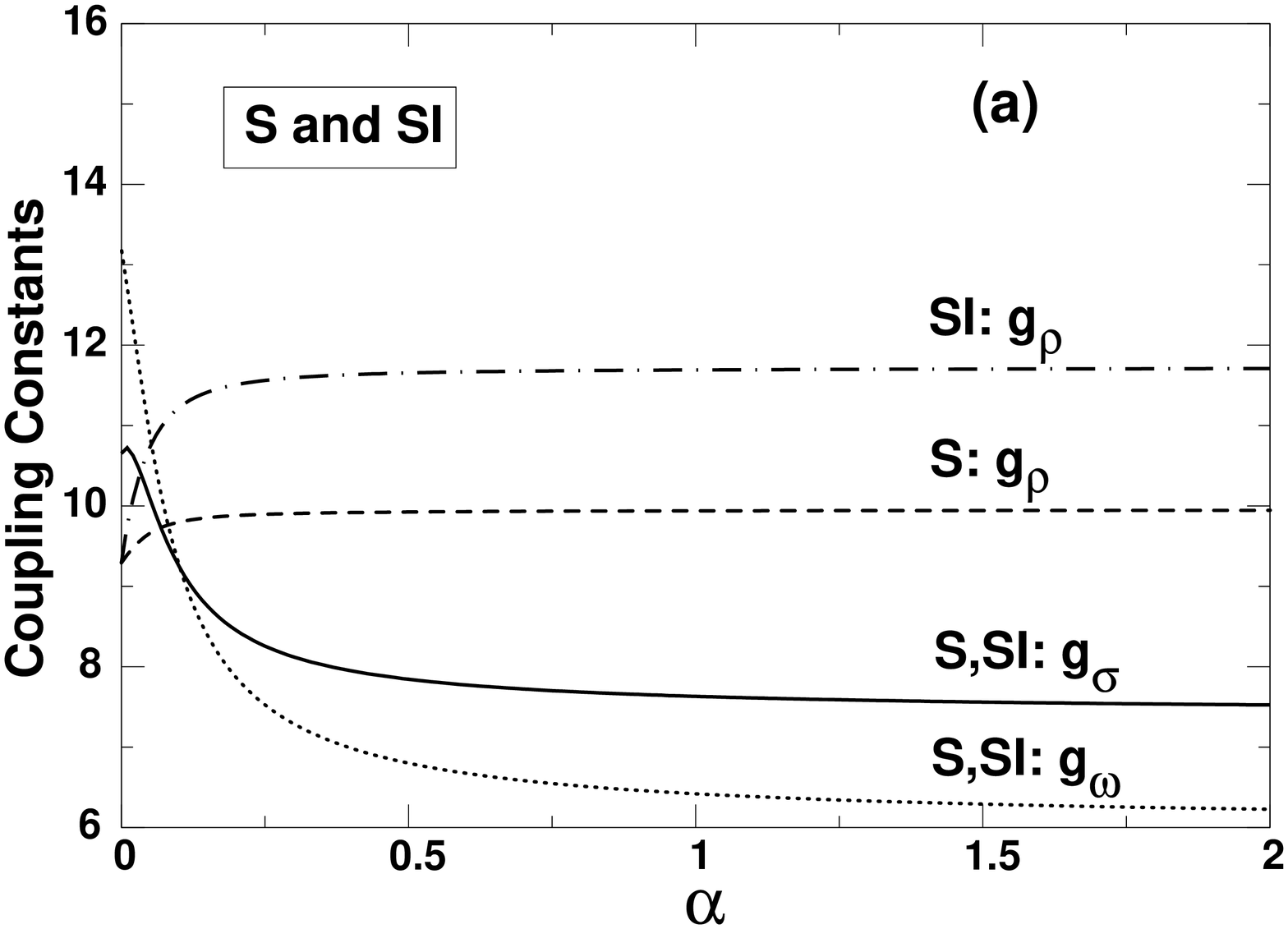}\\
$\phantom{ccccccccc}$\\
\includegraphics[width=0.8\linewidth]{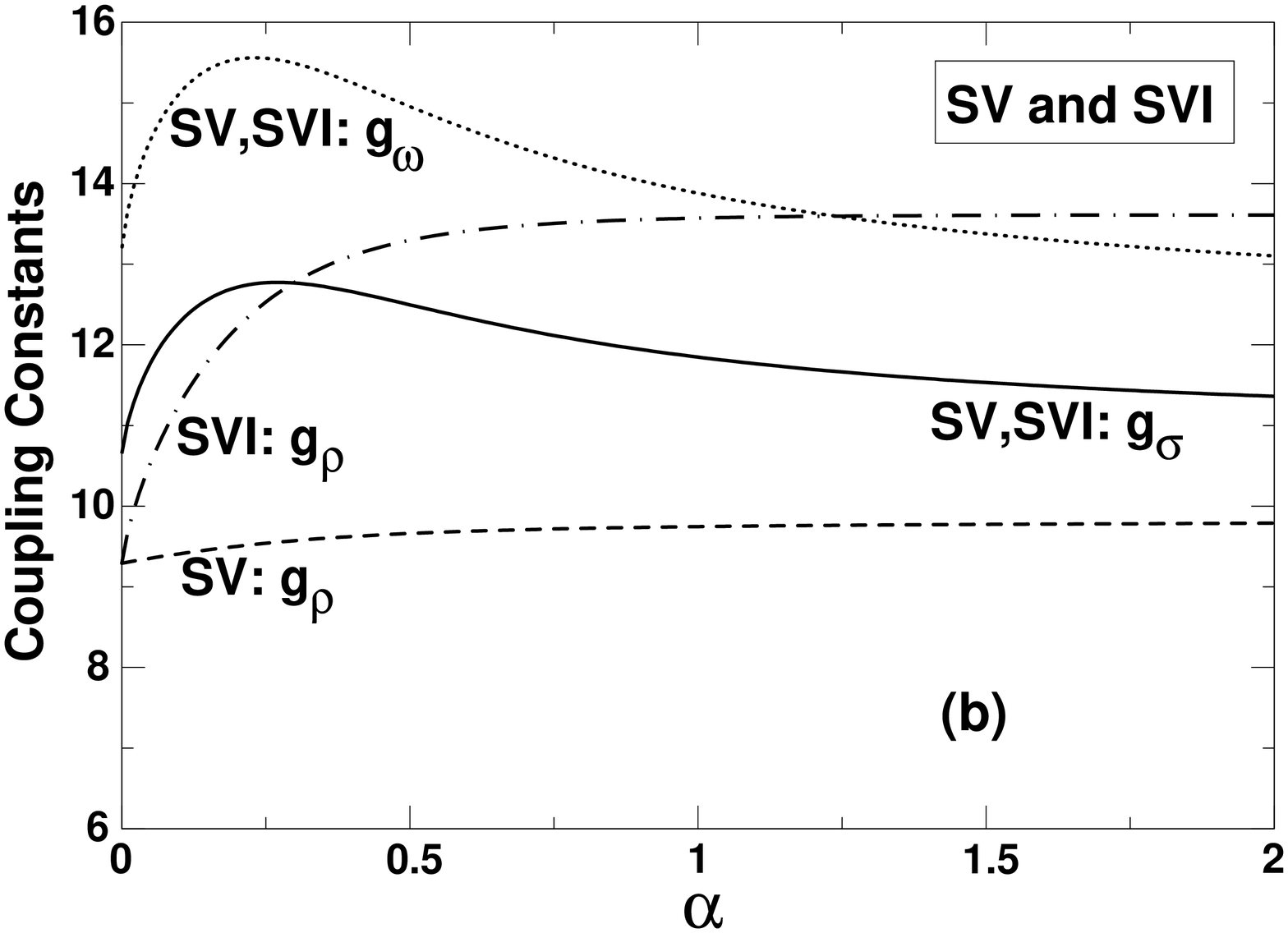}\\
\end{tabular}
\end{center}
\caption{The coupling parameters $g_i$ of the  models a) S and SI; b)
  SV and SVI,  as a function of $\alpha$. For each couple of models
  the $g_\sigma$ and $g_\omega$ parameters are equal and only the
  $g_\rho$ couplings vary. $g_\sigma$ (solid line), $g_\omega$ (dotted
  line), $g_\rho$ (dashed line for S and SV, dot-dashed line for SI.}
\label{coup}
\end{figure}

The parametric couplings restore the Walecka model for $\alpha=\beta=\gamma=0$, the ZM1 model for $\alpha=1$ and $\beta=\gamma=0$, the ZM3 model for $\alpha=\beta=1$ and $\gamma=0$, and exponential coupling model for $\alpha=\beta=\gamma=\infty$.

The $g_\sigma$ and $g_\omega$ coupling
constants are chosen to reproduce the binding energy
$E_B=\epsilon/\rho - M=-15.75$ MeV at the saturation density
$\rho_0=0.16$ fm$^{-3}$. We  fix the isovector coupling constant, $g_\rho$ to fit the symmetry energy, $a_4=32.5 $ MeV, at the same saturation density.
\begin{figure}[!]
\begin{center}
\begin{tabular}{c}
\includegraphics[width=0.95\linewidth]{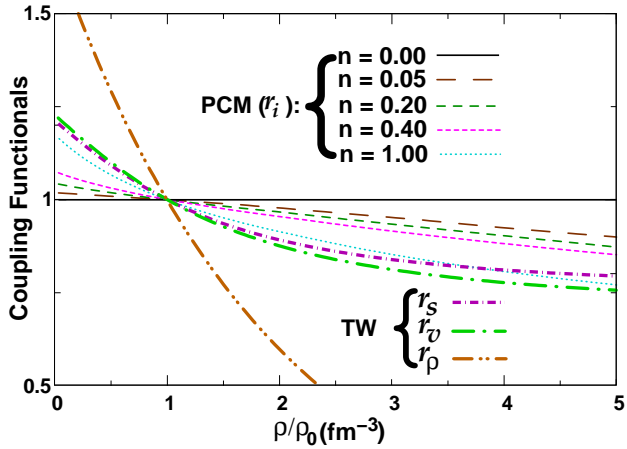}
\end{tabular}
\end{center}
\caption{(Color online) The effective coupling constant to bare coupling constant
ratio $g^*_i/g_i$ for $n$=0.0 (full), 0.05 (brown, long dashed), 0.20 (dark green, medium dashed), 0.40 (pink, short dashed) and 1.0 (cyan, dotted line) with $i=\sigma,\omega,\rho$ for different
choices of $\alpha$ in the PCM, as well as TW functionals $r_\sigma$ (purple, short dot-dashed), $r_\omega$ (light green, long dot-dashed) and $r_\rho$ (light brown, dot-dot-dashed line), where $r_i=\Gamma_i/\Gamma_0$. Notice that TW curves are shown with thicker lines.}
\label{g}
\end{figure}

We will work with four versions of the PCM: a) varying the scalar parameter $\alpha$ with the vector parameter $\beta=0$ and isovector parameter $\gamma=0$ (Model-S); b) varying the scalar and vector parameters
$\alpha=\beta$ while keeping the isovector parameter $\gamma=0$ (Model-SV);
c) varying all three parameters $\alpha=\beta=\gamma$ (Model-SVI) and d)
varying the scalar and isovector parameters $\alpha=\gamma$ while keeping the
vector parameter $\beta=0$ (Model-SI). We expect to verify the nuclear matter
main properties dependence on the coupling constants. We will also consider the parametrization a) SI2 with $\alpha=0.1$, $\beta=0$ and 
b) SVI2 $\alpha=\beta=0.2$ while, in both cases, we vary $\gamma$. These last two choices will allow us to explore the isovector degree of freedom while choosing reasonable properties for the isoscalar
channel, namely for symmetric nuclear matter.

In Fig. \ref{coup}, we represent the meson couplings $g_\sigma$, $g_\omega$ and 
$g_\rho$, as a function of the parameter $\alpha$ for all models. We plot the couplings for models S and SI in one figure and the ones of models SV and SVI in the other. Since the values of the  $g_\sigma$ and $g_\omega$ couplings are first chosen to reproduce symmetric nuclear matter data and only after this choice we establish the value of $g_\rho$, the scalar and vector coupling constants are the same for the pairs of models S, SI and SV, SVI. These two couples of models differ only in the isovector channel. It is seen that for models S and SI the $g_\rho$ parameter has stabilized at a constant value for $\alpha\sim 0.2$. For the models SV and SVI this occurs only for $\alpha\sim 0.5$. The $g_\sigma$ and $g_\omega$ parameters stabilize only for $\alpha\sim 1$, however, saturation properties of symmetric nuclear matter restrict the accepted values to  $\alpha<1$. In all four different versions of the model, $\alpha=\infty$ has a stable result which essentially does not differ from $\alpha=2$ and the changes from $\alpha=1$ are very small. The density dependence of the coupling constants can be examined from Fig.\ref{g}, where we have plotted $\frac{g^*_i}{g_i}=m_n^*=\left(1+\frac{g_\sigma\sigma}{nM_B}\right)^{-n}$.

The greater the value of $n$ the
stronger the density dependence and the reduction of the effective
coupling constant value at higher densities. In order to compare the
results obtained with PCM, we have also plotted, in the same figure,
the density dependence of the relative TW couplings \cite{tw}.
The behavior of $\Gamma_i/\Gamma_{i0}, \, i=\sigma,\, \omega$ is
similar to the one obtained with PCM for $\alpha=1$. The same does
not occur for $\Gamma_\rho$ which has a much faster decrease with density.

We have included hyperons in the present model for the  high density EOS in the inner regions of a neutron star. Since the discussion of the hyperon coupling constants is not the aim of this work, we just fix the hyperon-meson coupling constants to $\chi=\frac{g_{\sigma Y}}{g_{\sigma N}}=\frac{g_{\omega Y}}{g_{\omega N}}=\frac{g_{\rho Y}}{g_{\rho N}}=\sqrt{2/3}$. In order to study neutron stars we must also include the Baym-Pethick-Sutherland (BPS) EOS for densities below the neutron drip line \cite{BPS}.
\begin{figure}[!]
  \begin{center}
\begin{tabular}{c}
\includegraphics[width=0.9\linewidth]{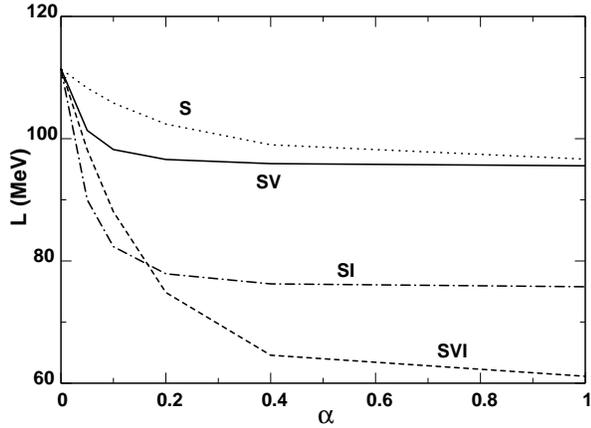}
\end{tabular}
\end{center}
\caption{The dependence of the slope of symmetry energy on the free parameter $\alpha$ for model S (dotted line), SV (solid line), SI (dot-dashed 
line) and SVI (dashed line).}
\label{ll}
\end{figure}

\section{Symmetry Energy \label{sec:esym}}

\begin{figure}[!]
\begin{center}
\begin{tabular}{cc}
\includegraphics[width=0.9\linewidth]{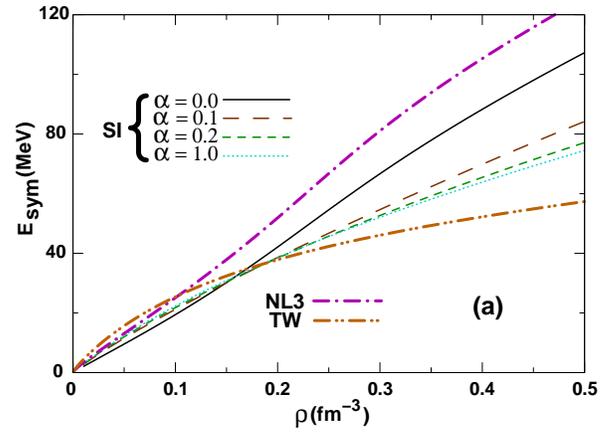}\\
$\phantom{mmmmmm}$\\
\includegraphics[width=0.9\linewidth]{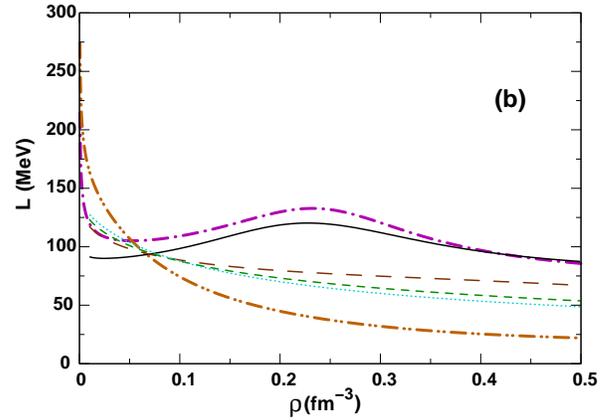}\\
$\phantom{mmmmmm}$\\
\includegraphics[width=0.9\linewidth]{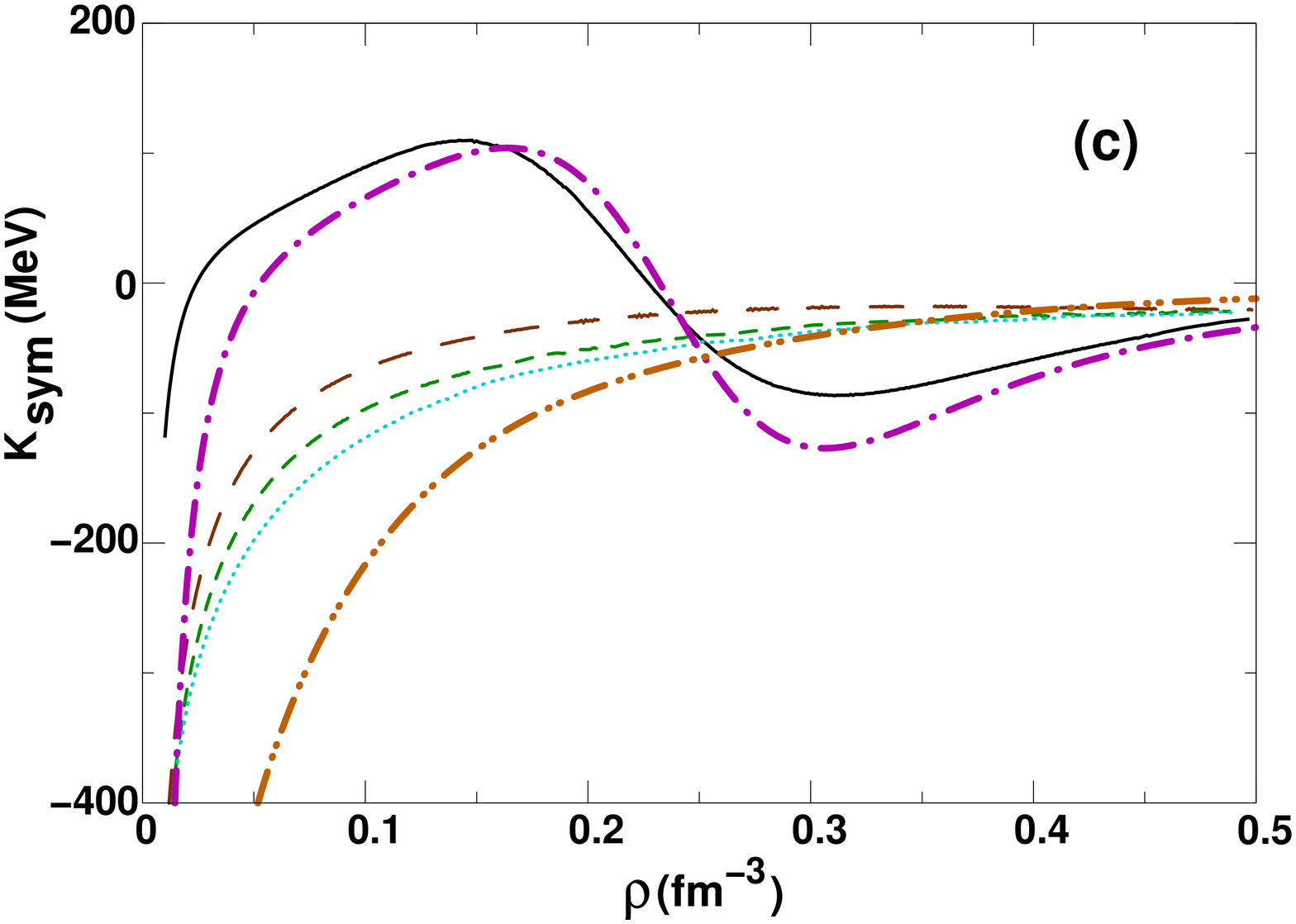}\\
\end{tabular}
\end{center}
\caption{(Color online) a) Symmetry energy, b) the symmetry energy  slope $L$ and c) the  incompressibility $K_{sym}$ for model SI with $\alpha$=0, 0.10, 0.20, 1.00=$\gamma$; $\beta$=0.
legend: $\alpha=0$ (full line), 0.1 (dark brown long dashed line), 0.2 (green short dashed line), 1.0 (cian dotted line), NL3 (purple dot-dashed line) and TW  (light brown dot-dot-dashed line).}
  \label{esym}
\end{figure}

\begin{figure*}[t]
\begin{center}
\begin{tabular}{c}
\includegraphics[width=0.7\linewidth]{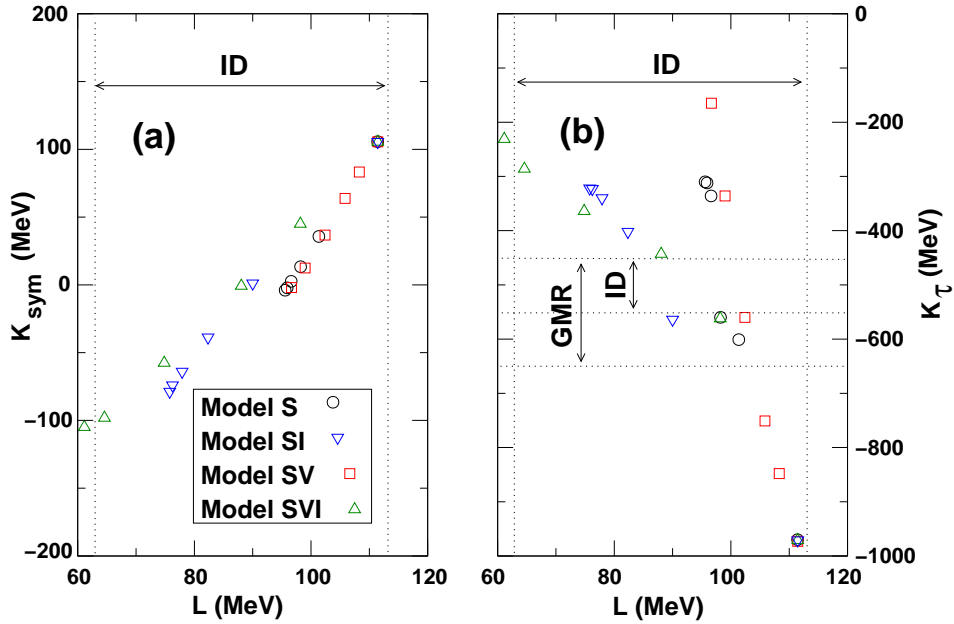}
\end{tabular}
\end{center}
\caption{(Color online) $K_{sym}$ and  $K_{\tau}$  as a function of  the symmetry energy  slope $L$ for all the parametrizations studied. The presently accepted  limits from  isospin diffusion (ID) in heavy-ion collisions \cite{chen05,li08}, giant monopole resonance (GMR)  in Sn isotopes \cite{garg07,li07} and neutron skin studies (NS) \cite{cente09} .}%
\label{exp}
\end{figure*}

In Tables \ref{tab1}-\ref{tab4}  we give some of the properties of the models under study: the symmetry energy slope $L=3\rho_0\partial E_{sym}/\partial\rho$, the symmetry energy incompressibility  $K_{sym}=9\rho_0^2\partial^2 E_{sym}/\partial\rho^2$, the symmetry term of the
incompressibility of the nuclear EOS  $K_\tau=K_{sym}-L(6-Q_0/K)$, where $Q_0=27\rho_0^3\partial^3 (E/A)/\partial \rho^3$, the effective nucleon mass $M^*/M$, the onset density of strangeness $\rho_s$, the onset density of muons $\rho_\mu$, the threshold density for the direct Urca process  $\rho_{DU}$ and the mass of a star with central density equal to $\rho_{DU}$, $M_{DU}/M_\odot$, the mass  $M_{MAX}/M_\odot$ and radius of the stable star configuration with maximum mass.

The slope $L$ is within the experimental constraints 55 MeV $<L< 113$ MeV for
all parametrizations considered. In the S and SV models the slope has a quite
large value, though never below 95 MeV. The variation of L as the free parameter
$\alpha$ increases is shown in Fig.\ref{ll}. It is seen that $L$ decreases
with the increase of $\alpha$, starting at $\sim 110$ MeV and stabilizing at
$\sim 95$ MeV for S and SV models,  at
$\sim 75$ MeV for the SI  model, and at $\sim 60$ MeV for SVI.

A value of $K_\tau$ in the interval [-400, -675] MeV puts very strong restrictions. According to some authors \cite{pieka09,piek07,sagawa07} however, it is difficult to determine the experimental value of $K_\tau$  accurately which  may suffer from the same ambiguities already encountered in earlier attempts to extract the incompressibility coefficient of infinite nuclear matter from finite-nuclei extrapolations. We, therefore, do not rule out models with values of $K_\tau$ larger than -400 MeV.
In fact, in \cite{isaac09} the properties of several relativistic and Skyrme models which reproduce the ground state properties of stable and unstable nuclei or the properties of nuclear or neutron matter  have been compared and it was shown that a large number of models have   $K_\tau< -400$ MeV.

\begin{table*}[th]
\caption{Results for Model-S. The binding energy (15.75 MeV), saturation density (0.16 fm$^{-3}$) and  symmetry energy (32.5 MeV) at saturation are the same for all parametrizations. The quantities given are the symmetry energy slope $L$, the symmetry energy incompressibility  $K_{sym}$, the symmetry term of the
incompressibility of the nuclear EOS  $K_\tau$, the effective nucleon mass at saturation density $M^*/M$, the onset density of strangeness $\rho_s$, the onset density of muons $\rho_\mu$, the threshold density for the direct Urca process  $\rho_{DU}$ and the mass of a star with central density equal to $\rho_{DU}$, $M_{DU}/M_\odot$, the mass  $M_{MAX}/M_\odot$ and radius of the stable star configuration with maximum mass.  }
\begin{tabular}{cccccccccccccc} \hline
$\alpha$ & $L$  & $K_{sym}$ & K  & $K_\tau$ & $Q_0$ 
& $M^*/M$& $\rho_t$&$\rho_s $ & $\rho_\mu$ & $\rho_{DU}$ & $M_{DU}$ 
& $M_{MAX}$  & R  \\
& (MeV)&  (MeV)&  (MeV)&  (MeV)& (MeV) & & $(\mbox{fm}^{-3})$& $(\mbox{fm}^{-3})$& $(\mbox{fm}^{-3})$& $(\mbox{fm}^{-3})$&($M_\odot$)&($M_\odot$)&(km)
\\ \hline
0.00 & 111.42& 105.67& 566 & -970 &-2068.27&0.537&0.1025&0.244&0.109 & 0.194 & 0.77 &2.58&12.61 \\
0.05 & 101.35& 35.74 & 310 & -601 &-87.60&0.650&0.0969&0.282&0.109&0.210&0.71&2.11&10.68\\
0.10 & 98.22 & 13.39 & 224 & -490 &195.97&0.737&0.0952&0.309&0.110&0.221&0.68&1.79&9.50\\
0.20 & 96.60 & 2.55  & 212 & -336 &529.01&0.798&0.0950&0.330&0.111&0.229&0.68&1.54&9.31\\
0.40 & 95.92 & -2.15 & 218 & -312 &603.79&0.833&0.0951&0.339&0.111&0.234&0.68&1.48&9.58\\
1.00 & 95.57 & -3.95 & 224 & -310 &626.67&0.850&0.0951&0.345&0.112&0.236&0.68&1.49&9.91\\ \hline
\end{tabular}
\label{tab1}
\end{table*}

\begin{table*}[th]
\caption{The same of Table \ref{tab1}  for Model-SI}
\begin{tabular}{cccccccccccccc} \hline
$\alpha$ & $L$  & $K_{sym}$ & K  & $K_\tau$ & $Q_0$ 
& $M^*/M$& $\rho_t$& $\rho_s $ & $\rho_\mu$ & $\rho_{DU}$ & $M_{DU}$ 
& $M_{MAX}$  & R  \\
& (MeV)&  (MeV)&  (MeV)&  (MeV)& (MeV) && $(\mbox{fm}^{-3})$& $(\mbox{fm}^{-3})$& $(\mbox{fm}^{-3})$& $(\mbox{fm}^{-3})$&($M_\odot$)&($M_\odot$)&(km)
\\ \hline
0.00 & 111.42& 105.67 & 566 & -970&-2068.27&0.537&0.1025&0.244& 0.109 & 0.194 & 0.77&2.58&12.61 \\
0.05 & 90.00 & 1.05 & 310 & -564   &-86.28&0.650&0.0970&0.284 & 0.107 & 0.213 & 0.69 &2.11&10.69\\
0.10 & 82.34 & -38.77 & 224 & -402 &355.86&0.737&0.0954& 0.314 & 0.107 & 0.229 & 0.65 &1.80&9.50\\
0.20 & 77.90 & -64.01 & 212 & -340 &520.91&0.798&0.0952& 0.339 & 0.107 & 0.248 & 0.65 &1.55&9.26\\
0.40 & 76.25 & -74.09 & 218 & -323 &596.36&0.833&0.0952& 0.354 & 0.106 & 0.264 & 0.67 &1.48&9.44\\
1.00 & 75.78 & -78.79 & 224 & -322 &625.09&0.850&0.0954& 0.366 & 0.106 & 0.278 & 0.69 &1.49&9.53\\ \hline
\end{tabular}
\label{tab2}
\end{table*}

\begin{table*}[th]
\caption{The same of Table \ref{tab1}  for Model-SV}
\begin{tabular}{cccccccccccccc} \hline
$\alpha$ & $L$  & $K_{sym}$ & K  & $K_\tau$ &$Q_0$
& $M^*/M$& $\rho_t$& $\rho_s $ & $\rho_\mu$ & $\rho_{DU}$ & $M_{DU}$ 
& $M_{MAX}$  & R  \\
& (MeV)&  (MeV)&  (MeV)&  (MeV)& (MeV)&& $(\mbox{fm}^{-3})$& $(\mbox{fm}^{-3})$& $(\mbox{fm}^{-3})$& $(\mbox{fm}^{-3})$&($M_\odot$)&($M_\odot$)&(km)
\\ \hline
0.00 & 111.42 & 105.67&566& -970 &-2068.27&0.537&0.1025&0.244 & 0.109& 0.194 & 0.77 &2.58&12.61 \\
0.05 & 108.29 & 83.26 & 458 & -848  &-1190.65&0.554&0.0938&0.254&0.108&0.197&0.77&2.42&11.96\\
0.10 & 105.84 & 63.78 & 360 & -751  &-611.36&0.570 &0.0861&0.263&0.108&0.199&0.73&2.26&11.31\\
0.20 & 102.40 & 36.71 & 276 & -560  &47.68&0.600   &0.0793&0.281&0.108&0.204&0.69&1.96&10.06\\
0.40 & 99.00 & 12.42 & 195 & -336   &483.72&0.649  &0.0727&0.310&0.108&0.212&0.63&1.39&8.70\\
1.00 & 96.66 & -1.99 & 159 & -165   &685.86&0.710  &0.0699&0.342&0.109&0.221&0.62&0.83&9.00\\ \hline
\end{tabular}
\label{tab3}
\end{table*}

\begin{table*}[th]
\caption{The same of Table \ref{tab1}  for Model-SVI}
\begin{tabular}{cccccccccccccc} \hline
$\alpha$ & $L$  & $K_{sym}$ & K  & $K_\tau$ &$Q_0$ 
& $M^*/M$& $\rho_t$& $\rho_s $ & $\rho_\mu$ & $\rho_{DU}$ & $M_{DU}$ 
& $M_{MAX}$  & R  \\
& (MeV)&  (MeV)&  (MeV)&  (MeV)& (MeV)& & $(\mbox{fm}^{-3})$& $(\mbox{fm}^{-3})$& $(\mbox{fm}^{-3})$& $(\mbox{fm}^{-3})$&($M_\odot$)&($M_\odot$)&(km)
\\ \hline
0.00 & 111.42 &105.67&566& -970 &-2068.27&0.537&0.1025&0.244 & 0.109 & 0.194 & 0.77 &2.58&12.61 \\
0.05 & 98.18 & 45.03 & 458 & -562   &-83.73&0.554&0.0940&0.254&0.106&0.198&0.75&2.42&11.96\\
0.10 & 88.05 & -0.74 & 360 & -443   &351.78&0.570&0.0863&0.265&0.105&0.203&0.68&2.26&11.31\\
0.20 & 74.84 & -57.56 & 276 & -364  &525.89&0.600&0.0795&0.286&0.102&0.214&0.64&1.97&10.07\\
0.40 & 64.58 & -98.07 & 195 & -286  &602.54&0.649&0.0731&0.323&0.100&0.239&0.59&1.40&8.65\\
1.00 & 61.13 & -104.91 & 159 & -231 &626.04&0.710&0.0704&0.374&0.098&0.293&0.57&0.66&8.67\\ \hline
\end{tabular}
\label{tab4}
\end{table*}

For reference, in Fig. \ref{esym} we plot for the different parametrizations
considered the symmetry energy (a), its slope $L$ (b) and the incompressibility $K_{sym}$ (c), for
model SI. We also include the corresponding curves obtained with the
non-linear Walecka model parametrization  NL3 \cite{nl3} and density dependent
relativistic hadronic model TW \cite{tw}. Varying $\alpha$ and $\gamma$
between 0 and 1, we go from the  hard symmetry energy behavior of NL3 to the
soft TW behavior. Similar conclusions are taken for the slope and the
incompressibility: as we increase $\alpha$ the behavior of our parametric model changes
from a NL3-like to a TW-like behavior.
\begin{figure}[!]
\begin{center}
\begin{tabular}{c}
\includegraphics[width=0.95\linewidth]{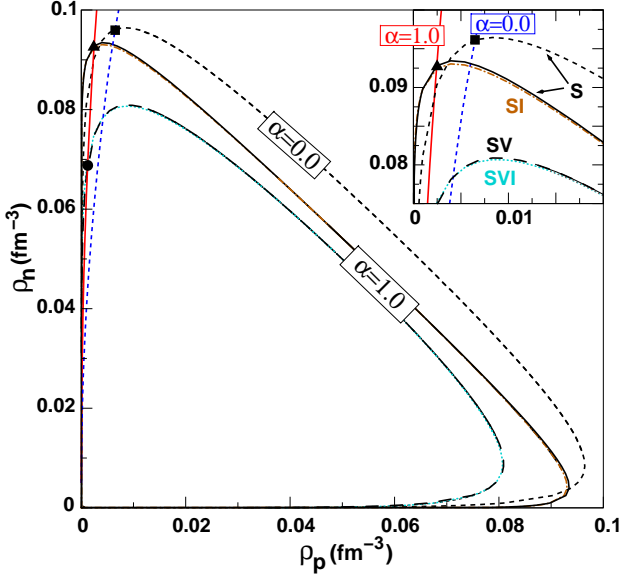}
\end{tabular}
\end{center}
\caption{(Color online) Spinodal curves for S, $\alpha=0$ (short dashed), S and SI,
$\alpha=1$ (full and brown dot-dashed), SV and SVI
$\alpha=1$ (long dashed and cyan, dotted lines), as indicated. The red, dot-dashed
and dark blue dotted lines represent the $\beta$-equilibrium EOS for 
$\alpha=\beta=\gamma=1$ (red) and $\alpha=\beta=\gamma=0$ (blue line), and the circle, triangle and square stress the crossing point of the EOS with the spinodal for their respective $\alpha$ values. The inset magnifies the crossing of the $\beta$-equilibrium EOS with the spinodal.}
\label{spinodal}
\end{figure}

In order to compare the predictions of the different parametrizations under study and the
experimental constraints on $L$ and $K_\tau$ we plot for all the models $K_{sym}$ and $K_\tau$
versus the slope $L$ in Fig. \ref{exp}. It is seen $K_{sym}$ and $K_\tau$  show a linear
correlation with $L$ as also discussed in \cite{isaac09}. We have checked that in fact this
correlation also exists for the ratio $Q_0/K$ which enters the definition of  $K_\tau$.
 We include the constraints obtained  from  isospin diffusion (ID) in heavy-ion collisions \cite{chen05,li08}, isotopic behavior of the  giant monopole resonance (GMR)  in Sn isotopes \cite{garg07,li07} and neutron skin studies (NS) \cite{cente09} .  The models which better satisfy the constraints are: S and SI, with $0.05<\alpha<0.1$ and SVI, with $0.05<\alpha<0.2$. Model SV only presents values with good agreement with $L$ and $K_\tau$ for $\alpha\sim 0.2$. So far we have considered the S, SV,  SI and SVI versions of the PCM in which two, one or none of the parameters has been set to zero, while the remaining ones have been set equal. In the next sections we still keep this analysis to impose some constraints on the parameter choice, using neutron stars observations. We finally vary the parameters independently and choose two sets which give excellent values  not only for $L$ and $K_\tau$ but also for $K$ and neutron star maximum masses.

\section{Thermodynamical Instabilities\label{sec:thermo}}

The system is unstable against phase separation keeping volume and temperature
constant, if the free energy  curvature matrix ${\cal F}_{ij}$
\begin{equation}
{\cal F}_{ij}=\left( \frac{\partial^{2} {\cal F}}{\partial \rho_{i}\partial\rho_{j}}\right) _{T}
\end{equation}
 is negative \cite {chomaz03}. Stability implies that the free energy density $\cal F$ is a convex function of the densities $\rho_p$ and $\rho_n$. In the present section, we define the free energy $ {\cal F}$ with no leptons or hyperons.

Since we are interested in studying the spinodal curve of the two-fluid nuclear system, it is enough to evaluate the zero of the smallest  eigenvalue $\lambda_-$ of ${\cal F}_{ij}$ , where $$\lambda_-=\frac{1}{2}\left(\mbox{Tr}({\cal F})\pm\sqrt{\mbox{Tr}({\cal F})^2-4\mbox{Det}({\cal F})}\right),$$
and where the relevant eigenvectors are $$\frac{{\delta\rho_j}^-}{{\delta\rho_i}^-}=\frac{\lambda_--{\cal F}_{ii}}{{\cal F}_{ij}},$$
with $i,j=p,n$ \cite{chomaz03,Avancini-06}.

The thermodynamic spinodal, defined by the $\lambda_-=0$ condition determines
the instability regions of the system and is plotted in Fig. \ref{spinodal}
for the models under discussion. We point out that, for symmetric matter, all
the models have the same  density, binding energy and symmetry energy at
saturation but different incompressibilities and effective nucleon masses. The
model with the largest incompressibility (S, $\alpha=0$) has the largest
spinodal and the ones with the smallest incompressibility (SV and SVI,
$\alpha=1$) have the smallest spinodal. The point of the spinodal with
$\rho_p=\rho_n$ corresponds to the minimum of the pressure of symmetric nuclear matter, when the incompressibility is zero.  Since we impose that the pressure is zero for all models at the same density, the saturation density is 0.16 fm$^{-3}$, the incompressibility of the models with a larger curvature will become zero at a larger density. We notice that the effect of the parameter $\gamma$ is very small and only seen for very asymmetric matter, when it slightly reduces the spinodal.

In Fig. \ref{spinodal} we also plot the $\beta$-equilibrium EOS for $\alpha=0$ and for model
SVI with $\alpha=1$.
The crossing of the $\beta$-equilibrium EOS with the thermodynamical spinodal instability line
gives a prediction for the transition density approximately 15\% larger than the value obtained from a Thomas-Fermi calculation of the pasta phase \cite{pasta}. Therefore, we 
 determine the crust-core transition density from the crossing between the EOS and the
spinodal and expect that our estimation of the transition density  will define
an upper bound of the correct transition density \cite{pethick95} between the
crust and the core of a neutron star. It is seen that quite different
predictions are obtained within PCM for the  transition density both because the spinodals
have a different behavior for neutron rich matter and the EOS for
$\beta$-equilibrium stellar matter also differ when varying $\alpha$ from 0 to
1. Even if the EOS do not differ considerably for different
parameters, the values of the density and isospin asymmetries at the crossing of the EOS with the
spinodal are found to be quite different, though always with low proton densities. 
In Fig. \ref{spinodal} we only plot the EOS for $\alpha=0$ and $\alpha=1$ for SVI. We
have obtained for the transition densities:   $0.102$ fm$^{-3}$ ($\alpha=0$), $\rho_t\sim 0.095$ fm$^{-3}$ (models S and SI with $\alpha=1$)  and  $\rho_t\sim 0.070$
fm$^{-3}$ (models SV and SVI with $\alpha=1$). 
The values of $\rho_t$ given in Tables \ref{tab1}$\sim$\ref{tab4} show a decrease of $\rho_t$ with the decrease of the slope $L$, contrary to the results of \cite{isaac09}. However, we have to
take these values with care because we are not only varying the slope $L$ but also the
incompressibility $K$ while imposing the same saturation density for all models. As a
consequence, the extension of the spinodal region grows with $K$ and it is this correlation that
is reflected in the data of Tables \ref{tab1}-\ref{tab4}.
In the next section we calculate the transition density for a selected set of parameters and
then we will be able to obtain a  correlation between $L$ and $\rho_t$ similar to the one obtained in \cite{isaac09}.

\section{Neutron Star Constraints\label{sec:ns}}

\begin{figure}[!]
\begin{center}
\begin{tabular}{c}
\includegraphics[width=0.95\linewidth]{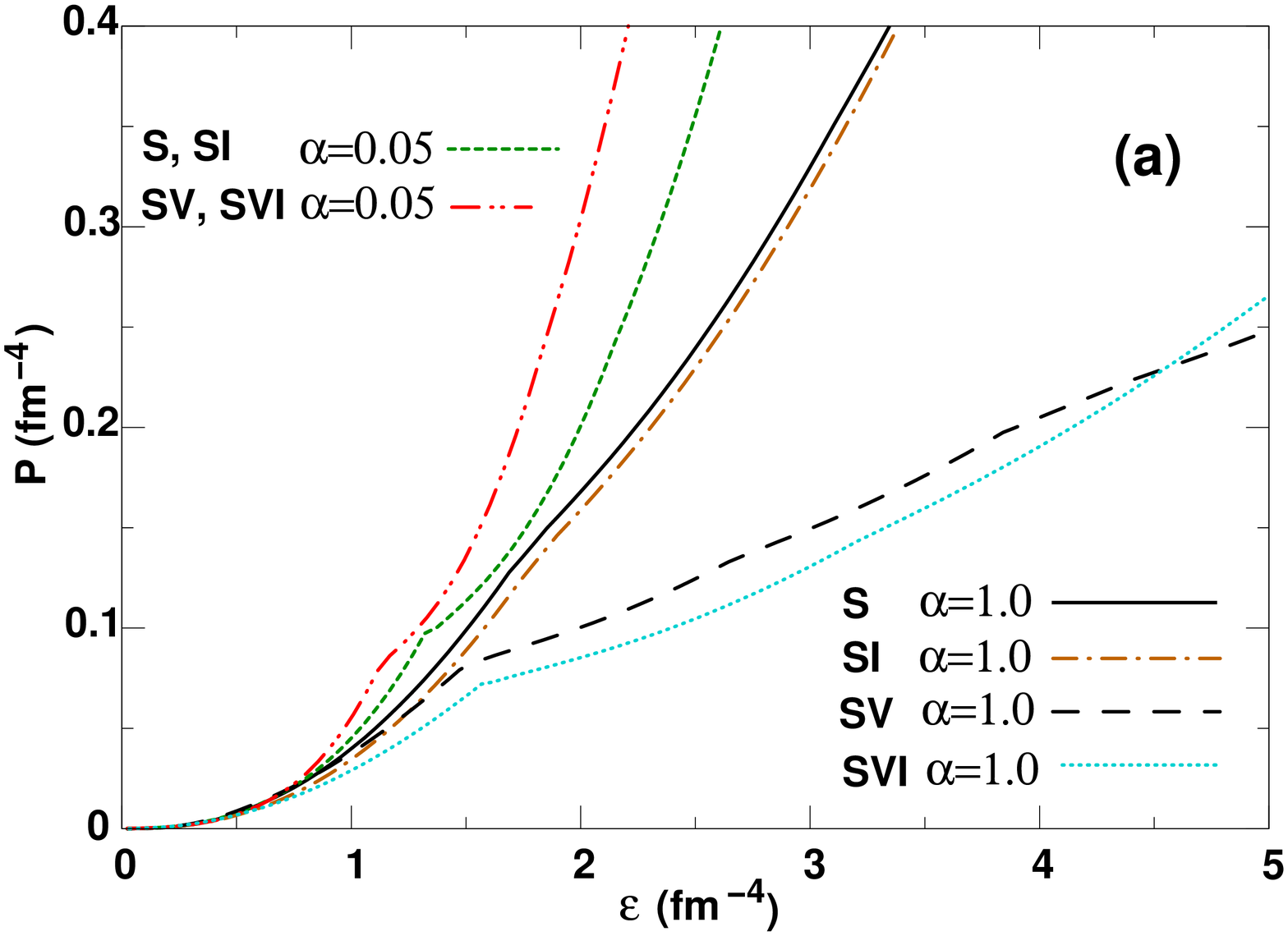}\\
$\phantom{mmmmmmm}$\\
\includegraphics[width=0.95\linewidth]{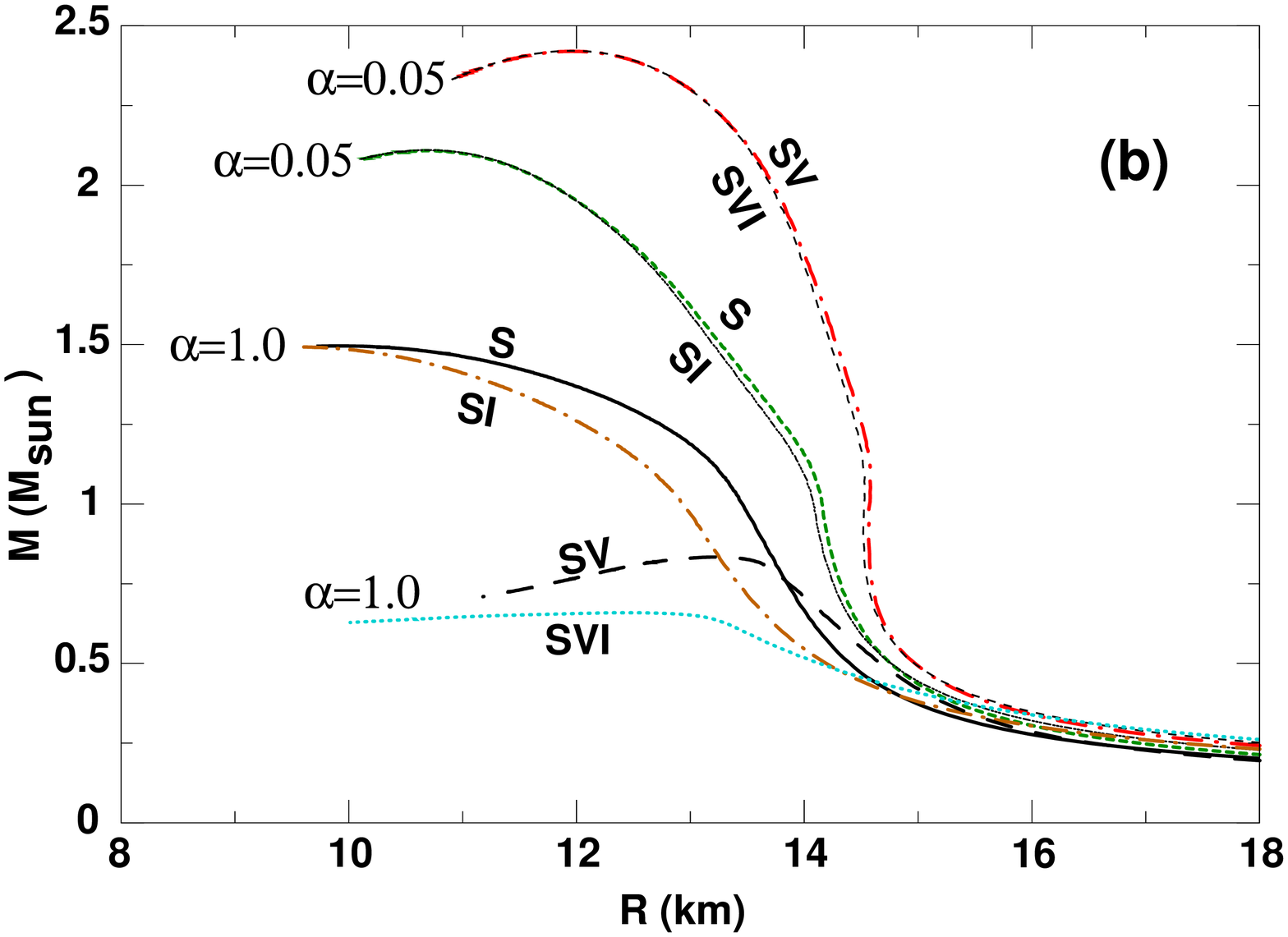}
\end{tabular}
\end{center}
\caption{(Color online) a) Equation of state and b) it corresponding Mass - Radius curves for the families of stars obtained with the parametrizations of Tables \ref{tab1}-\ref{tab4}. As indicated on the figure, for $\alpha$=0.05 the coincident S and SI curves are shown by the green, short dashed line, whereas the red, dot-dot-dashed line stands for the SV, SVI single curve. For $\alpha$=1.0, the S, SI, SV, SVI curves are shown by full, brown dot-dashed, long dashed and cyan dotted lines, respectively.}
\label{eos}
\end{figure}

Besides heavy ion experiments, we can make use of astrophysical observations of neutron stars to stablish some constraints on the PCM and its free parameters. We therefore obtain the EOS of $\beta$-equilibrium matter, including the presence of hyperons, and integrate the Tolman-Oppenheimer-Volkoff equation for a spherical compact star in hydrostatic equilibrium \cite{TOV1,TOV2} in order to obtain the family of stars which correspond to each parametrization considered.

The different EOS are shown in Fig.\ref{eos}a)  for all four versions of the model, considering the limiting values $\alpha=0.05$ and $\alpha=1.00$. The results for values of $0.05<\alpha<1.00$ are intermediate and lie between the two curves. The kink on each curve corresponds to the onset of strangeness. For $\alpha=0.05$ the incompressibility is 310 MeV for S and SI models and 458 MeV for SV and SVI. The last EOS is the hardest one and therefore, the onset of hyperons occurs at smaller densities for this model. The hyperons make the EOS softer but never softer than any of the other versions (with higher values of $\alpha$) of the model.

The corresponding neutron star families (mass-radius) are shown in Fig.\ref{eos}b). As expected, one can see that all models with low values of $\alpha$ have maximum neutron star masses above 2$M_\odot$, while the results drastically change when values of $\alpha$ approach the unit, reducing the maximum neutron star mass obtained with the S and SI models to values close to 1.5$M_\odot$. The results get even smaller when SV and SVI model are considered, reducing the maximum neutron star mass to values below 1$M_\odot$.  

One can stablish the minimum value required for the maximum neutron star mass from
observations. From Tables \ref{tab1}-\ref{tab4} we  see that, for the S and SI models, which present almost the same neutron star masses, $\alpha<0.1$ allow neutron star masses greater the 1.8$M_\odot$, while the models SV and SVI have a wider range of values ($\alpha<0.3$) that yield the same star mass.

\begin{figure}[b]
\begin{center}
\begin{tabular}{c}
\includegraphics[width=0.95\linewidth]{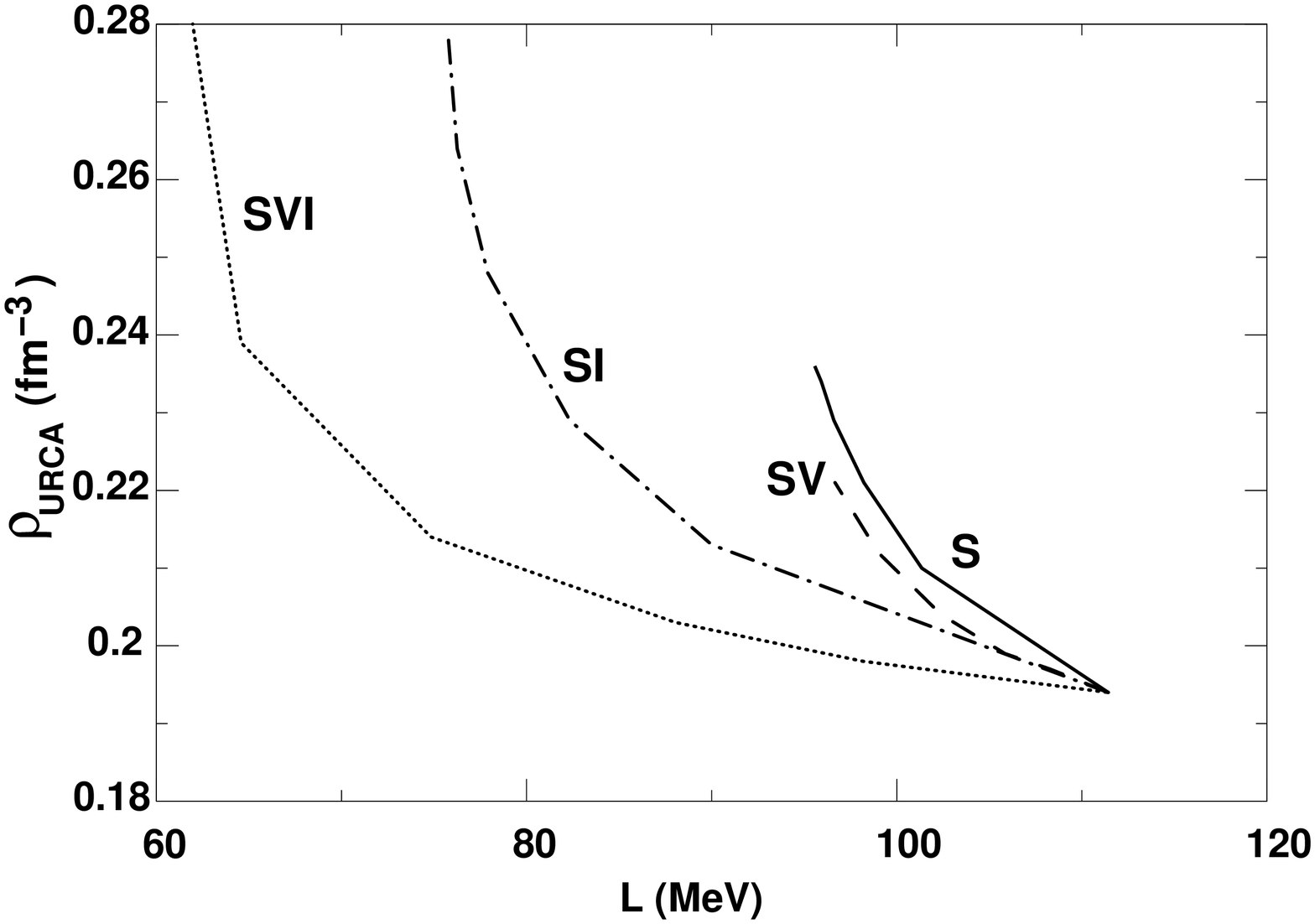}
\end{tabular}
\end{center}
\caption{Threshold density for the direct Urca process to occur as versus the slope $L$. The cyan dotted, brown dot-dashed, dashed and full lines stand for the SVI, SI, SV and S models, respectively.}
\label{urca}
\end{figure}

According to \cite{klan06}, observations on cooling predict that direct Urca should occur only in stars with a mass $M> 1.35 M_\odot$. 
We have determined  the star with maximum mass that does not allow the direct Urca process ($M_{DU}$) for the different parametrizations of 
the model proposed. By $\rho_{DU}$ we denote the baryonic density that defines the onset of the
Urca process. The results are summarized in Tables \ref{tab1}-\ref{tab4}. No  model satisfies the condition  $M_{DU}> 1.35 M_\odot$. All
four versions of the model predict the Urca process with densities 0.194
fm$^{-3}<\rho_{DU}<0.293 \mbox{ fm}^{-3}$ which correspond to neutron stars masses
around $0.57\, M_\odot<M_{DU}<0.77\, M_\odot$. This result is expected as PCM
presents the same main features as the models discussed in
\cite{klan06} which are not in agreement with Urca expectations.
The threshold density for the direct Urca process is plotted versus 
the slope $L$ in Fig. \ref{urca}. It is seen that it strongly 
depends on the slope of the symmetry energy: a smaller slope corresponds to a larger density of the Urca onset.
\begin{table*}[th]
\caption{Results for $\alpha=0.1$ and $\beta=0$}
\begin{tabular}{ccccccccccccccc} \hline
$\gamma$ & $M^*/M$& $K$& $L $ & $K_{sym}$ &   $K_\tau$&$Q_0$ 
& $\rho_t $ & $\rho_s $ & $\rho_\mu $ & $\rho_{DU} $ & $M_{DU}$ 
& $M_{MAX}$ & R  \\
& &(MeV)&(MeV)&(MeV)&(MeV)&(MeV)&(fm$^{-3}$) &(fm$^{-3}$) &(fm$^{-3}$)&(fm$^{-3}$)&(M$_\odot$) & (M$_\odot$) &(Km)
\\ \hline
0.450 &0.749 & 224 &60 & -123.5 & -585 & -378.93&0.0894& 0.305 & 0.107 & 0.253 & 0.68 & 1.811 & 9.512 \\
0.240 & 0.749& 224 &70 & -87.7  & -587 & -253.76&0.0874 &0.305 & 0.107 & 0.240 & 0.67 & 1.806 & 9.510 \\
0.120 & 0.749 & 224&80 & -46.3  & -583 & -158.76&0.0865 &0.305 & 0.107 & 0.231 & 0.66 & 1.803 & 9.511 \\ \hline
\end{tabular}
\label{tab5}
\end{table*}

\begin{table*}[th]
\caption{Results for $\alpha=\beta=0.2$}
\begin{tabular}{ccccccccccccccc} \hline
$\gamma$ & $M^*/M$& $K$& $L $ & $K_{sym}$ &   $K_\tau$&$Q_0$ 
& $\rho_t $ & $\rho_s $ & $\rho_\mu $ & $\rho_{DU} $ & $M_{DU}$ 
& $M_{MAX}$ & R  \\
& &(MeV)&(MeV)&(MeV)&(MeV)&(MeV)&(fm$^{-3}$) &(fm$^{-3}$) &(fm$^{-3}$)&(fm$^{-3}$)&(M$_\odot$) & (M$_\odot$) &(Km)
\\ \hline
0.310&0.60& 276  & 60 & -82.2 &-517 & -344.08& 0.0884 &0.286 & 0.102 & 0.218 & 0.65 &1.971&10.072\\
0.210 &0.60& 276 & 70 & -57.3 & -539 &-243.27& 0.0877 &0.286 & 0.102 & 0.214 & 0.64 &1.968&10.077\\
0.125  &0.60&276 & 80 & -28.5 & -547 &-132.82& 0.0868 &0.286 & 0.102 & 0.210 & 0.63 &1.966&10.069\\ \hline
\end{tabular}
\label{tab6}
\end{table*}

\begin{figure*}[!]
  \begin{center}
\begin{tabular}{c}
    \includegraphics[width=12cm]{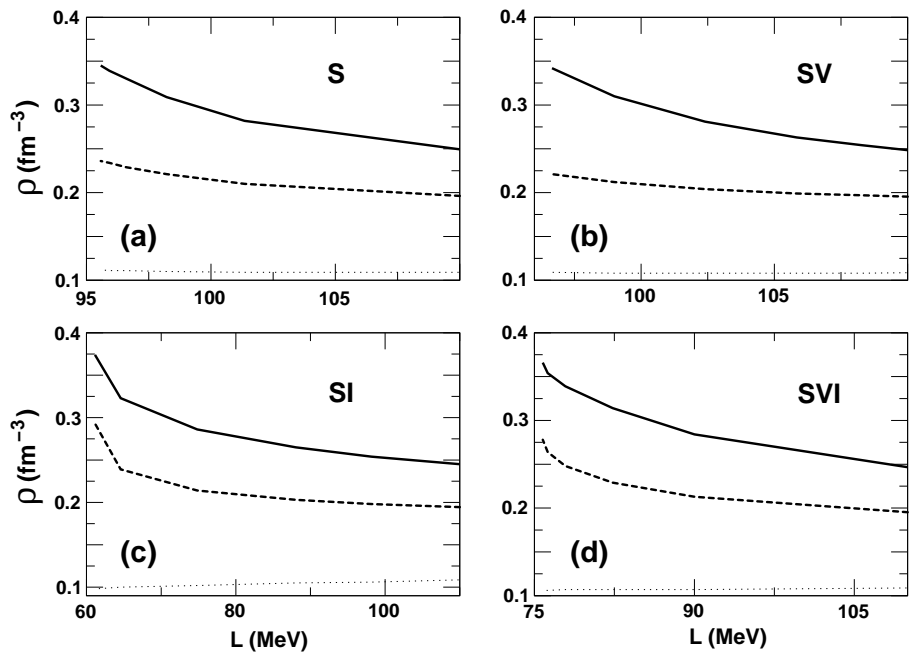}
\end{tabular}
\end{center}
\caption{Density for the strangeness (full), the Urca process (dotted) and the muon (dashed line) onset for each version of the model.}
  \label{lvarios}
\end{figure*}
We expect to be able to infer isospin asymmetry properties or stablish a few constraints on the parametrizations from 
neutron star observations. When comparing S and SI models or SV and SVI models, which present the same global behavior, we have identified a
small difference in the neutron star radii that becomes larger as $\alpha$ increases. For the S and SI models the difference between the radii reaches $\Delta R$=380 m, for $\alpha=1.0$ while it reaches $\Delta R$=330 m for SV and SVI model. This difference would correspond, for example, to a difference of $\Delta \nu_f\sim$ 0.1 kHz ($\sim 4\% $) in the fundamental mode of a star \cite{benhar,araujo} and is larger if we consider stars with masses below the maximum allowed mass.

In Fig.\ref{lvarios} several properties of the neutron star structure are plotted as a function of the symmetry energy slope for the models S, SV, SI and SVI. The figure  shows the baryonic densities at the onset of hyperons for each model. Hyperons are very important because their appearance results in the softening of the EOS and the lowering of the neutron star maximum mass. In the same figure, it is also represented the baryonic density at the muon onset and the density at which the proton fraction reaches the value which allows the direct  Urca process to occur in neutron stars (for details see Ref.\cite{urca,klan06}). While the muon onset is not very sensitive to the value of $L$, the hyperon onset may vary 40-50\%, and the densities for the onset of Urca, 20-50\% between the limits of the plotted values of $L$ which are all within the experimental constrained values.

\begin{figure*}[!]
  \begin{center}
\begin{tabular}{c}
\includegraphics[width=0.9\linewidth]{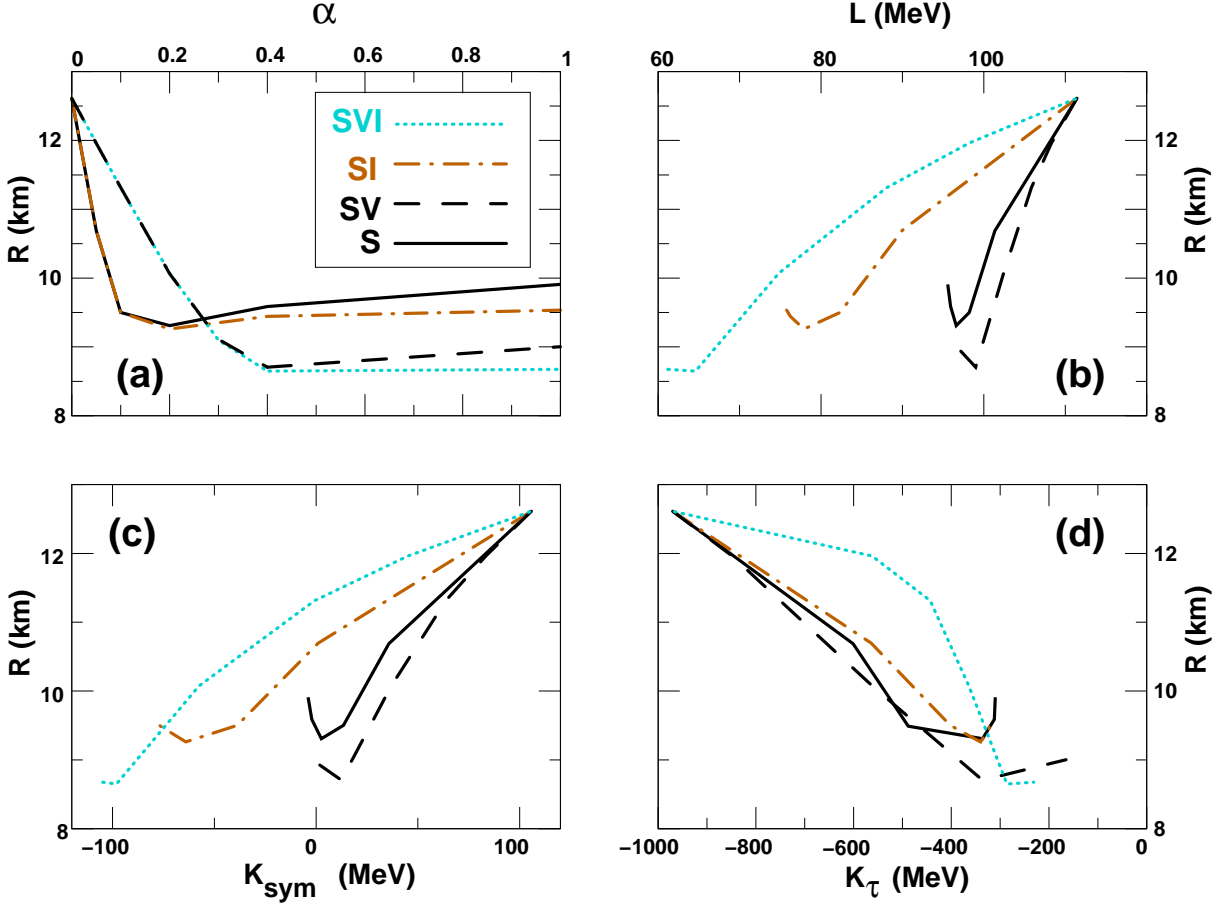}
\end{tabular}
\end{center}
\caption{(Color online) Radius of star with maximum mass versus $\alpha$, $L$, $K_{sym}$ and $K_\tau$, for the four different versions of the PCM: S (full), SI (brown dot-dashed), SV (dashed) and SVI (cyan dotted line).}
\label{radius}
\end{figure*}

In Fig.\ref{radius} we analyse  the dependence of the maximum mass neutron star radius on $\alpha$ and the 
asymmetry parameters $L$, $K_{sym}$ and $K_\tau$. It is shown that there
is a correlation which may help to impose restrictions on the parameters: smaller values of
$L$, $K_{sym}$ and $|K_\tau|$ give larger radius for the maximum mass configuration.

We propose to use the new parametrizations SI2 and SVI2 because models S, SI, SV and
SVI with reasonable incompressibilities predict too small maximum star masses.
With a fixed $\alpha$ and/or $\alpha=\beta$ parameter, and changing the $\gamma$
parameter we have changed the symmetry energy slope. Tables \ref{tab5} and
\ref{tab6} show the properties of the EOS and neutron stars for these new
parametrizations. The maximum star masses have improved and are of the order
1.8 $M_\odot$ while the properties of symmetric nuclear matter are all
reasonable. However the density for Urca onset is still too small in all
models and the radius of the maximum mass configuration is also quite small.
We have tried to vary all the parameters independently of each other but
  we were not able to find a parametrization which predicts larger densities
  for the onset of the Urca process than the ones already obtained. This seems
to be a weak point of these models.
We made also an estimation of the crust-core  transition
densities by calculating the crossing of the $\beta$-equilibrium EOS with the thermodynamical
spinodal as explained in section IV. The results are given in Tables  \ref{tab5} and \ref{tab6} and vary between 0.086 and
0.089 fm$^{-3}$ showing  a small decrease when the slope of the symmetry energy increases.

We next  compare our results with two  RMF models: NL3 with constant
couplings and TW with density dependent coupling parameters (Table
\ref{tw_nl3}). We have obtained for the models SI2 and SVI2 better
values for  the slope of the symmetry energy, well within the experimental range,  than NL3  and  TW give. For these two models the
slope of the symmetry energy  corresponds, 
respectively, to the upper and lower limit of $L$ proposed from the experimental values.
 We have also  obtained results for $K_\tau$ closer to the experimental ones than those found within
NL3 and TW, keeping good results for the compression modulus, effective nucleon mass and
neutron star global properties. PCM introduces a
density dependence on the coupling parameters weaker than the one of TW (see Fig. \ref{g}),
mainly for the isospin channel. The  parametrization of the coupling parameters
within PCM does not allow for such a strong density dependence for the coupling
parameters  as TW shows.
 For the crust-core transition density the models SI2 and SVI2 preview a density similar to the
one obtained with TW. Concerning the correlation between  $\rho_t$ and $L$, models SI2 and
SVI2 behave like those in Ref.\cite{isaac09}, decreasing the value of $\rho_t$ as $L$
increases.
The density for the strangeness onset is closely related with the
incompressibility of the EOS: SI2 is very soft and predicts the largest  density for the
strangeness onset, SVI2 gives a result very similar to TW, $\sim 40\%$ larger than NL3. Other
properties of SI2 and SVI2 are also closer to TW than to NL3 except for the the onset density
for the Urca process and the  mass of the compact star with this density at the centre: it
is seen that for other similar saturation properties, the value of $K_\tau$ for TW seems to be the only great difference.

\begin{table*}[th]
\caption{Results for TW and NL3}
\begin{tabular}{cccccccccccccccc} \hline
Model & $B/A$ & $~\rho_0$ & $K$  & $M^*/M$ & ${\mathcal E}_{sym}$& $L$ & $K_{sym}$ &  $K_\tau$  
& $\rho_t$ & $\rho_s$ & $\rho_\mu$ & $\rho_{DU}$ & $M_{DU}$ 
& $M_{MAX}$ & R \\ 
& (MeV)  &(fm$^{-3}$)& (MeV) & & (MeV) & (MeV) & (MeV) & (MeV) &(fm$^{-3}$) &(fm$^{-3}$)&(fm$^{-3}$)&(fm$^{-3}$)&(M$_\odot$) & (M$_\odot$) &(Km)\\
\hline
TW & 16.3  & 0.153     & 240.1& 0.56    & 32.0              &55.3 & -124.7 & -332.1& 0.085&0.287 & 0.115 & 0.315 & 1.46 & 2.012 & 11.60 \\
NL3  & 16.3  & 0.148     & 270  & 0.60    & 37.4        & 118.5 & 100.9 & -698.4 &0.065& 0.217 & 0.111 & 0.205 & 1.00 & 1.707 & 12.65 \\
\hline
\end{tabular}
\label{tw_nl3}
\end{table*}

\section{Conclusions\label{sec:concl}}
We have employed the parametric coupling model, developed by Taurines
{\it et. al.} \cite{taurines}, to describe nuclear matter  properties, in special,
those related to the isospin channel which were still not investigated
in the PCM
model. Some of the parameters of the model were fixed by the binding
energy
(15.75 MeV) and  saturation density  (0.16 fm$^{-3}$) for symmetric
matter and
by the symmetry energy (32.5 MeV). We have tried to restrict the remaining
parameters by  reproducing experimental results obtained with heavy ion
collisions at intermediate energies and neutron star properties.

Considering a wide range of experimental data  for the slope of the
symmetry energy $L$,  55 MeV $<L<115$ MeV and for  the symmetry term 
of the incompressibility of the nuclear EOS $K_\tau$,
$-375$ MeV$ <K_\tau<-650$ MeV, we have ruled out some values of the
free parameters. The models which best satisfy the constraints are: S and SI, with
$0.05<\alpha\lesssim 0.1$ and SVI, with $0.05<\alpha\lesssim 0.2$.
Model SV only presents values with good agreement with the  $L$ and $K_\tau$ experimental
values  for $0.2\lesssim \alpha\lesssim 0.3 $.

Some properties of neutron star matter and neutron star structure have
been described with the PCM for the range of the parameters
restricted by the values of $L$ and $K_\tau$, namely the maximum mass star
configuration, and the densities of the onset of muons, strangeness and the direct
Urca process. It was shown that all properties were reasonably well described except the
prediction of the direct Urca process at a too low density. It was
also shown that the density at the onset of muons is almost independent of the
parameters of the model but the strangeness onset is sensitive to the symmetry energy
slope: a larger slope corresponds to a smaller onset density. The same
is true for the direct Urca process, although with a slightly weaker
dependence. It was also shown that the radius and mass of the maximum mass
stable star configuration was correlated with the $L$ value: a larger $L$ corresponds to a larger
radius and a larger mass. A  similar correlation exists with the symmetry term of the
incompressibility of nuclear matter. However from model to model this correlation has a different slope and a given value of $L$ and $K_\tau$  does not define in a unique way the neutron star properties.

Even though the proposed range of validity of the parameters of the
PCM are in agreement with the experimental range described
above for heavy ions, it is of common belief that the slope of
symmetry energy should be close to the lower limit of the interval 55
MeV$<L<115$ MeV
\cite{cente09}.
In order to investigate other  possibilities of the model, we have
proposed new parametrizations chosen to reproduce the symmetry energy
slope $ L$=60, 70 and 80 MeV. These parametrizations have shown excellent results also for
$K_\tau$, $K$, $M^*$ and the neutron star maximum mass but still predict quite
low densities for the onset of the Urca process. The results shown in tables
\ref{tab5} and \ref{tab6} describe some of these results. We have tried to
find a set of parameters which could predict a larger density for the
onset of the direct Urca process without success. The PCM
present the same main features of the models discussed in
\cite{klan06} which are not in agreement with Urca expectations.

We have constrained the free parameters of the PCM using the properties of asymmetric nuclear matter, but it is important to mention that the EoS which show the best results (tables V and VI), for these properties are in perfect agreement with those imposed by the analysis of symmetric nuclear matter properties such as those described by \cite{dan02} where constraints, obtained from the analysis of nuclear matter flow in heavy ion collisions, were proposed for the high density EoS of symmetric matter. This problem has been recently examined in Ref.\cite{tobepub}. It is also important to mention that the constraints imposed by asymmetric nuclear matter properties are stronger than the ones obtained from symmetric nuclear matter (for details see Ref.\cite{tobepub}).

We have  compared our results with  two  RMF models: NL3 with constant
couplings and TW with density dependent coupling parameters (Table
\ref{tw_nl3}). It was shown that the models SI2 and SVI2 predict results closer to TW except for
the onset density of the Urca process. These could be related to the larger $K_{\tau}$ value
that TW has since all the other saturation properties are similar. It was also shown that
while the density dependence of the $g^*_{\sigma}$ and $g^*_{\omega}$ coupling parameters in
PCM is similar to the one of TW, the density dependence of $g^*_{\rho}$ is very different.

The relation between the symmetry energy
and some properties of neutron stars was also discussed. The
different models considered in the present study with one or more density dependent couplings, showed quite large
differences among them, even for the same values of  $L$ and $K_\tau$. We conclude
that the properties of nuclear matter for both symmetric and asymmetric matter at
saturation are not enough to define the structure of compact stars and
experimental data at two, three times saturation density are required.
\section{ACKNOWLEDGMENTS}
This work was partially supported by FEDER and FCT under the projects PTDC/FP/64707/2006 and CERN/FP/83505/2008,  and SFRH/BPD/29057/2006 and  by COMPSTAR, an ESF Research Networking Programme.


\vspace{0.1cm}

\end{document}